\documentclass[letterpaper,american,twocolumn,showpacs,pre,aps,superscriptaddress]{revtex4}
\usepackage[latin1]{inputenc}
\usepackage{bm}
\usepackage{multirow,amssymb,amsbsy,amsmath}
\usepackage{amsmath}
\usepackage{graphicx}
\usepackage{amssymb}
\makeatletter
\usepackage{pifont}
\makeatother

\newcommand{\bignone}{}
\newcommand{\mathd}{\mathrm{d}}

\newcommand{\tmop}[1]{\ensuremath{\operatorname{#1}}}

\begin{document}

\title{Exact master equations for the non-Markovian decay of a
qubit}

\author{Bassano Vacchini}

\email{vacchini@mi.infn.it}

\affiliation{Universit{\`a} degli Studi di Milano, Dipartimento di
Fisica, Via Celoria 16, I-20133 Milano, Italy } \affiliation{INFN
Sezione di Milano, Via Celoria 16, I-20133 Milano, Italy}

\author{Heinz-Peter Breuer}

\email{breuer@physik.uni-freiburg.de}

\affiliation{Physikalisches Institut, Universit\"at Freiburg,
             Hermann-Herder-Strasse 3, D-79104 Freiburg, Germany}

\date{\today}

\begin{abstract}
Exact master equations describing the decay of a two-state system
into a structured reservoir are constructed. Employing the exact
solution for the model we determine analytical expressions for
the memory kernel of the Nakajima-Zwanzig master equation and for
the generator of the corresponding time-convolutionless master
equation. This approach allows a detailed investigation and
comparison of the convergence behavior of the corresponding
perturbation expansions. Moreover, we find that the structure of
widely used phenomenological master equations with memory kernel
may be incompatible with a non-perturbative treatment of the
underlying microscopic model. We discuss several physical
implications of our results on the microscopic analysis and the
phenomenological modelling of non-Markovian quantum dynamics of
open systems.
\end{abstract}


\maketitle

\section{Introduction}\label{sec:introduction}
The field of open quantum systems
\cite{Alicki2007,Breuer2007,Weiss2008} is of great interest
because of its relevance in applications of quantum mechanics, as
well as in a deeper understanding of the theory itself. Indeed the
study of the interaction between a quantum system and its
environment is an endeavor common to many fields such as quantum
measurement theory, quantum communication, quantum optics,
condensed matter theory and quantum chemistry to name a few. The
field is well assessed as far as Markovian dynamics is concerned,
in which the Gorini-Kossakowski-Sudarshan-Lindblad expression for
the generator of a quantum dynamical semigroup
\cite{Gorini1976a,Lindblad1976a} provides a benchmark result for
both microscopic and phenomenological approaches. This situation
is however not satisfactory when one has to go beyond the
Born-Markov approximation and considers systems in which a
separation of time scales between system and environment can no
longer be assumed in a realistic description. Memory effects then
become important and a non-Markovian description is mandatory. For
this case a general consistent theoretical framework has not yet
been found, and partial results have been obtained as a result of
intense efforts (see, e.g.,
Refs.~\cite{Budini2004a,Budini2005a,Budini2005b,Budini2006a,Breuer2006a,Breuer2007a,Krovi2007a,Vacchini2008a,Piilo2008a,Ferraro2008a,Breuer2008a,Kossakowski2008a,Kossakowski2009a,Breuer2009a,Chruscinski-xxx_1,Chruscinski-xxx}).
An important step in the development of a general theory consists
in the construction of a suitable measure that quantifies the
degree of non-Markovianity for a given dynamical evolution
\cite{Breuer2009b,Rivas-xxx}.

In this article we will obtain the exact Nakajima-Zwanzig kernel
for a two-level system coupled to a Bosonic reservoir discussed in
\cite{Breuer2007}, and compare it to the exact
time-convolutionless master equation as well as to the Markovian
approximation of the dynamics. This will show how involved the
transition from the approximate Markovian level of description to
the exact non-Markovian regime can be. Indeed, the non-Markovian
memory kernel is found to have an operator structure which differs
from the one that appears in the Born-Markov approximation. Often
one tries to obtain dynamical equations of motion for
non-Markovian systems by slight modifications with respect to the
Markovian case, e.g., by considering a master equation which
involves a superoperator given by a convolution in time of the
corresponding Markovian superoperator
\cite{Barnett2001a,Daffer2004a,Shabani2005a,Maniscalco2006a,Maniscalco2007a,Wilkie2009a}.
Our results show that such an approach, although being justified
as a phenomenological modelling, can be incompatible with a
non-perturbative treatment of the underlying microscopic
system-environment model. Moreover, different perturbation
expansions such as time-convolutionless and Nakajima-Zwanzig
projection operator technique turn out to have different ranges of
validity. Indeed the time-convolutionless expansion breaks down at
finite time in the strong coupling limit, while the
Nakajima-Zwanzig approach does not preserve positivity if
restricted to second order. Furthermore the convergence to the
exact solution is not uniform with respect to the expansion
parameter: Different matrix elements of the statistical operator
such as coherences and populations are obtained with quite
different accuracy at the same perturbative order.

The paper is organized as follows. In Sec.~\ref{MODEL} we
introduce the model and its exact solution, which is later
exploited to obtain the exact equations of motion for the reduced
statistical operator of the system. In
Sec.~\ref{sec:time-conv-mast} we recall the structure of the
time-convolutionless master equation, pointing out two different
perturbation expansions for the generator. In
Sec.~\ref{sec:nakaj-zwanz-mast} we derive the Nakajima-Zwanzig
integral kernel, providing an alternative expansion with respect
to the standard method. The two results are compared in
Sec.~\ref{sec:comp-time-local}, also building on an exact analytic
expression for all the quantities involved obtained considering a
Lorentzian spectral density. We finally draw our conclusions in
Sec.~\ref{sec:disc-concl}.

\section{The model and its exact solution}\label{MODEL}
The total Hamiltonian of the model is given by
\begin{equation} \label{HTOTAL}
 H= H_S + H_E + H_I = H_0 + H_I,
\end{equation}
where
\begin{equation} \label{H_0}
  H_S = \omega_{0}\sigma_+\sigma_-
\end{equation}
describes a two-state system (qubit) with ground state
$|0\rangle$, excited state $|1\rangle$ and transition frequency
$\omega_0$. The operators $\sigma_+ = |1\rangle\langle 0|$ and
$\sigma_- = |0\rangle\langle 1|$ are the raising and lowering
operators of the qubit. The environmental Hamiltonian is taken to
be
\begin{equation} \label{H_E}
 H_E = \sum_k\omega_k b_k^\dagger b_k,
\end{equation}
describing a collection of harmonic oscillators with creation and
annihilation operators $b_k^{\dagger}$ and $b_k$ which satisfy
bosonic commutation relations $[b_k,b_{k'}^{\dagger}] =
\delta_{kk'}$. The interaction Hamiltonian takes the form
\begin{equation} \label{H_I}
  H_I = \sum_k \left( g_k \sigma_+ \otimes b_k + g^*_k \sigma_- \otimes b^{\dagger}_k
  \right).
\end{equation}
The model thus describes for example the coupling of the qubit to
a reservoir of electromagnetic field modes labelled by the index
$k$ with corresponding frequencies $\omega_k$ and coupling
constants $g_k$, and has already been discussed in \cite{Breuer2007}.

In the following we will work in the interaction picture with
respect to $H_0 = H_S + H_E$. In this picture the
Schr\"odinger equation reads
\begin{equation} \label{SCHROEDINGER}
 \frac{d}{dt}|\Psi(t)\rangle = -iH_I(t)|\Psi(t)\rangle,
\end{equation}
where the interaction Hamiltonian is given by
\begin{equation}
\label{eq:hi}
 H_I(t) = \sigma_+(t) \otimes B(t) + \sigma_-(t) \otimes B^{\dagger}(t)
\end{equation}
with
\begin{equation}
 \sigma_{\pm}(t) = \sigma_{\pm} e^{\pm i \omega_0 t}
\end{equation}
and
\begin{equation}
 B(t) = \sum_k g_k b_k e^{-i \omega_k t}.
\end{equation}
It is easy to verify that the operator
\begin{equation}
 N = \sigma_+\sigma_- + \sum_k b_k^{\dagger}b_k
\end{equation}
for the number of excitations in the system commutes both with the
total Hamiltonian $H$ and with the interaction Hamiltonian
$H_I(t)$, which is a consequence of the fact that the rotating
wave approximation has been used in the interaction Hamiltonian
\eqref{H_I}. It follows that any initial state of the form
\begin{eqnarray} \label{PSI-INIT}
 |\Psi(0)\rangle &=& c_0 |0\rangle\otimes|0\rangle_E
 + c_1(0) |1\rangle\otimes|0\rangle_E  \\
 && + \sum_k c_k(0) |0\rangle\otimes|k\rangle_E \nonumber
\end{eqnarray}
evolves after time $t$ into the state
\begin{eqnarray} \label{PSI-T}
 |\Psi(t)\rangle &=& c_0 |0\rangle\otimes|0\rangle_E
 + c_1(t) |1\rangle\otimes|0\rangle_E  \\
 && + \sum_k c_k(t) |0\rangle\otimes|k\rangle_E. \nonumber
\end{eqnarray}
The state $|0\rangle_E$ denotes the vacuum state of the reservoir,
and $|k\rangle_E = b_k^\dagger|0\rangle_E$ the state with one
particle in mode $k$. Note that the amplitudes $c_1(t)$ and
$c_k(t)$ depend on time, while the amplitude $c_0$ is constant in
time because of $H_I(t)|0\rangle\otimes|0\rangle_E = 0$.
Substituting Eq.~\eqref{PSI-T} into the Schr\"odinger equation
\eqref{SCHROEDINGER} one finds
\begin{eqnarray}
 \frac{d}{dt}c_1(t) & = & - i \sum_k g_k e^{i(\omega_0-\omega_k)t} c_k(t),
 \label{EQ-1} \\
 \frac{d}{dt}c_k(t) & = & - i g_k^{\ast} e^{-i(\omega_0-\omega_k)t} c_1(t).
 \label{EQ-2}
\end{eqnarray}
We assume in the following that $c_k(0)=0$. This means that the
environment is in the vacuum state initially and that the total
initial state is given by the product state
\begin{equation} \label{INIT-PROD}
  |\Psi(0)\rangle = \big(c_0|0\rangle + c_1(0)|1\rangle\big)
  \otimes |0\rangle_E \equiv |\psi(0)\rangle \otimes |0\rangle_E.
\end{equation}
Expressing $c_k(t)$ in terms of $c_1(t)$ by means of
Eq.~\eqref{EQ-2} and substituting the result into Eq.~\eqref{EQ-1}
one obtains an integrodifferential equation for the amplitude
$c_1(t)$,
\begin{equation} \label{ExactC1Dgl}
 \frac{d}{dt}c_1(t) = - \int_0^t dt_1 f(t - t_1) c_1(t_1).
\end{equation}
Given the solution of this equation, which can be found through a
Laplace transformation, the amplitudes $c_k(t)$ are determined by
Eq.~\eqref{EQ-2}. The kernel $f(t-t_1)$ of Eq.~\eqref{ExactC1Dgl}
is given by a certain two-point correlation function of the
reservoir,
\begin{eqnarray}
\label{eq:7}
 f(t-t_1) &=& \langle 0|B(t)B^{\dagger}(t_1)|0\rangle e^{i\omega_0(t-t_1)}
 \\
 &=& \sum_k |g_k|^2 e^{i(\omega_0-\omega_k)(t-t_1)}, \nonumber
\end{eqnarray}
on which no restrictive hypothesis is made, so that our results will
be valid for an environment with a generic spectral density.

With the help of the procedure described above, already used by
Weisskopf and Wigner in their classical paper \cite{Weisskopf1930a}, one finds the
solution of the Schr\"odinger equation of the total system with
initial states of the form \eqref{INIT-PROD} lying in the sector
of the Hilbert space corresponding to zero or one excitations. By
means of this solution we can construct the exact dynamical map
describing the time-evolution of the reduced density matrix of the
qubit which is given by
\begin{equation} \label{RHO-QUBIT}
 \rho(t) = \mathrm{tr}_E \{ |\Psi(t) \rangle \langle \Psi(t)| \} =
 \left(\begin{array}{cc}
  \rho_{11}(t) & \rho_{10}(t) \\
  \rho_{01}(t) & \rho_{00}(t)
  \end{array}\right),
\end{equation}
where $\rho_{ij}(t)=\langle i|\rho(t)|j\rangle$ for $i,j=0,1$.
Using Eq.~\eqref{PSI-T} we find
\begin{eqnarray}
 \rho_{11}(t) &=& 1-\rho_{00}(t) = |c_1(t)|^2, \\
 \rho_{10}(t) &=& \rho_{01}^*(t) = c_0^*c_1(t).
\end{eqnarray}
It is convenient to define the function $G(t)$ as the solution of
the equation
\begin{equation} \label{G-DEF}
 \frac{d}{dt}G(t) = -\int_0^t dt_1 f(t-t_1)G(t_1)
\end{equation}
corresponding to the initial condition $G(0)=1$. We then have
$c_1(t)=G(t)c_1(0)$ and, hence, the dynamics of the elements of
the reduced density matrix can be represented as follows,
\begin{eqnarray}
 \rho_{11}(t) &=& |G(t)|^2 \rho_{11}(0), \label{DYN-MAP-1} \\
 \rho_{00}(t) &=& \rho_{00}(0) + (1-|G(t)|^2)\rho_{11}(0), \label{DYN-MAP-2} \\
 \rho_{10}(t) &=& G(t) \rho_{10}(0), \label{DYN-MAP-3} \\
 \rho_{01}(t) &=& G^*(t) \rho_{01}(0). \label{DYN-MAP-4}
\end{eqnarray}
These equations have been derived for the pure product initial
state \eqref{INIT-PROD}, i.e., they describe the time-evolution
corresponding to the pure reduced system's initial state
$\rho(0)=|\psi(0)\rangle\langle\psi(0)|$. However, since any mixed
initial state can be represented as convex-linear combination of
pure initial states, and since the function $G(t)$ introduced
above does not depend on the initial condition, the equations
\eqref{DYN-MAP-1}-\eqref{DYN-MAP-4} hold true for any pure or
mixed initial state. They thus represent the exact dynamical map
$\Phi(t)$ which transforms the initial states into the states at
time $t$:
\begin{equation}
 \Phi(t): \; \rho(0) \mapsto \rho(t) = \Phi(t)\rho(0), \qquad
 t\geq 0.
\end{equation}
Since we have constructed this map from the exact solution of the
model, it is clear from the general theory of open quantum systems
that $\Phi(t)$ is completely positive and trace preserving.

\section{The time-convolutionless master equation}\label{sec:time-conv-mast}

\subsection{Exact master equation in time-convolutionless form}\label{sec:exact-mast-equat}
The exact solution determined in Sec.~\ref{MODEL} enables the
construction of the exact generator ${\mathcal{K}}_{\rm TCL}$ of
the time-convolutionless master equation
\begin{equation} \label{QMEQ}
 \frac{d}{dt} \rho(t) = \mathcal{K}_{\rm TCL}(t) \rho(t)
\end{equation}
governing the dynamics of the reduced density matrix. The
time-convolutionless
generator is defined in terms of the dynamical map $\Phi(t)$ by
means of
\begin{equation}
\mathcal{K}_{\rm TCL}(t) = \dot{\Phi}(t)\Phi^{-1}(t)
\end{equation}
provided the inverse map $\Phi^{-1}(t)$ exists. Using then
Eqs.~\eqref{DYN-MAP-1}-\eqref{DYN-MAP-4} one shows that the generator takes the
following form \cite{Breuer2007},
\begin{eqnarray}
\label{TCL-GEN}
 \mathcal{K}_{\rm TCL}(t)\rho &=& -\frac{i}{2}S(t)
 [\sigma_+\sigma_-,\rho] \\
 &~& +\gamma(t)\left[ \sigma_-\rho\sigma_+
 -\frac{1}{2}\left\{\sigma_+\sigma_-,\rho\right\} \right],\nonumber
\end{eqnarray}
where we have introduced the definitions
\begin{equation} \label{eq:4}
 \gamma(t) = -2\Re\left(\frac{\dot{G}(t)}{G(t)}\right), \qquad
 S(t) = -2\Im\left(\frac{\dot{G}(t)}{G(t)}\right).
\end{equation}
By construction, Eq.~\eqref{QMEQ} with the generator
\eqref{TCL-GEN} represents an exact time-local master equation.
Note that the generator is well-defined as long as $G(t)\neq 0$.
The quantity $S(t)$ plays the role of a time-dependent frequency
shift, and $\gamma(t)$ can be interpreted as a time-dependent
decay rate. We observe that the structure of $\mathcal{K}_{\rm
TCL}$ is similar to that of a Lindblad generator. However, due to
the time dependence of the coefficients $S(t)$ and $\gamma(t)$
Eq.~\eqref{QMEQ} does generally not yield a quantum dynamical
semigroup. Moreover, the time-dependent rate $\gamma(t)$ may
become negative, signifying strong non-Markovian behavior of the
reduced system dynamics.

\subsection{Perturbation expansions of the generator}
In most cases of interest the time-convolutionless generator can
only be determined through a perturbation expansion. Here we
investigate two methods of expanding the exact master equation
\eqref{QMEQ} with respect to the strength of the interaction
Hamiltonian $H_I$. To this end, we introduce a small overall
expansion parameter $\alpha$, replacing the coupling constants
$g_k$ in the interaction Hamiltonian \eqref{H_I} by $\alpha g_k$.
The two-point correlation function $f(t)$, being proportional to
$\alpha^2$, is then to be regarded as a quantity of second order.

The first method consists in using
Eq.~\eqref{G-DEF} to obtain a perturbative expression for $G(t)$ from which one
directly finds an expansion for the coefficients $\gamma(t)$ and
$S(t)$ appearing in the master equation. The expansion of $G(t)$ is
obviously of the form
\begin{equation}
\label{expansion-G}
G(t) = \sum_{n=0}^{\infty} \alpha^{2n} G^{(2n)}(t),
\end{equation}
where $G^{(0)}(t)\equiv 1$ because of the required initial
condition $G(0)=1$, and Eq.~\eqref{G-DEF} leads to the following
recursion relation
\begin{equation}
\label{eq:6}
 G^{(2n)}(t) = -\int_0^t dt_1\int_0^{t_1} dt_2 f(t_1-t_2)G^{(2n-2)}(t_2).
\end{equation}
To illustrate the procedure we determine the frequency shift and
the decay rate to fourth order in $\alpha$:
\begin{eqnarray}
 \lefteqn{-\frac{1}{2}\left[\gamma(t)+iS(t)\right] =
 \frac{\dot{G}(t)}{G(t)}} \\
 && = \alpha^2\dot{G}^{(2)}(t) + \alpha^4
\left[\dot{G}^{(4)}(t)-\dot{G}^{(2)}(t)G^{(2)}(t)\right] +
 {\mathcal{O}}(\alpha^6). \nonumber
\end{eqnarray}
With the help of these expressions one obtains the second and the
fourth order contributions for the coefficients of the master
equation:
\begin{eqnarray}\label{eq:44}
 \gamma^{(2)}(t)+iS^{(2)}(t) &=& 2\int_0^t dt_1 f(t-t_1), \\
 \gamma^{(4)}(t)+iS^{(4)}(t) &=& 2
 \int_0^t dt_1 \int_0^{t_1} dt_2 \int_0^{t_2} dt_3 \nonumber \\
 && \!\!\!\!\!\!\!\!\!\!\!\!\!\!\!\!\!\!\!\!\!\!\! \times
 \left[f(t-t_2)f(t_1-t_3)+f(t-t_3)f(t_1-t_2)\right]. \nonumber
\end{eqnarray}

Another possibility for the construction of the perturbation
expansion is to use the general method of expanding the
time-convolutionless generator in terms of the ordered cumulants.
This procedure allows to write a closed expression for the
coefficients of the master equation which takes the form (for
details, see \cite{Breuer2007} and references therein):
\begin{eqnarray}
\label{cumulants}
  \gamma^{(2n)}(t) + i S^{(2n)}(t)  \! \!\!&=& \!\!\!\! \int_0^t dt_1
 \int_0^{t_1} dt_2 \ldots  \!
 \int_0^{t_{2n-2}} dt_{2n-1}  \\
 && \hspace{-30mm} \times 2 (-1)^{n+1} \langle f(t-t_1) f(t_2 - t_3)
 \ldots f(t_{2n-2} - t_{2n-1})
 \rangle_{\mathrm{oc}}. \nonumber
\end{eqnarray}

\section{The Nakajima-Zwanzig master equation}\label{sec:nakaj-zwanz-mast}
\subsection{The exact memory kernel}\label{sec:exact-memory-kernel}
The Nakajima-Zwanzig master equation is given by
\begin{equation} \label{NZ-MASTEREQ}
 \frac{d}{dt} \rho(t) = \int_0^t dt_1
 {\mathcal{K}}_{\rm NZ}(t-t_1) \rho(t_1),
\end{equation}
where the superoperator ${\mathcal{K}}_{\rm NZ}(\tau)$ represents
the memory kernel. We construct the form of this kernel from the
exact solution of our model obtained in Sec.~\ref{MODEL}. To this
end, we employ the following Ansatz,
\begin{eqnarray} \label{NZ-kernel}
 {\mathcal{K}}_{\rm NZ}(\tau) \rho &=&
 -i\varepsilon(\tau)[\sigma_+\sigma_-,\rho] \\
 &~& + k_1(\tau) \left[ \sigma_-\rho\sigma_+
 - \frac{1}{2} \left\{ \sigma_+\sigma_-,\rho \right\} \right] \nonumber \\
 &~& + k_2(\tau) \left[ \sigma_+\sigma_-\rho\sigma_+\sigma_-
 - \frac{1}{2} \left\{ \sigma_+\sigma_-,\rho \right\}
 \right], \nonumber
\end{eqnarray}
where the functions $\varepsilon(\tau)$, $k_1(\tau)$ and
$k_2(\tau)$ are real, such that the master equation preserves
Hermiticity and trace. The equations of motion for the population
$\rho_{11}(t)$ and the coherence $\rho_{10}(t)$ obtained from this
master equation read:
\begin{equation}  \label{rho11}
 \frac{d}{dt}\rho_{11}(t) = -\int_0^t dt_1 k_1(t-t_1)
 \rho_{11}(t_1),
\end{equation}
and
\begin{eqnarray} \label{rho10}
 \lefteqn{\frac{d}{dt}\rho_{10}(t) = -\int_0^t dt_1 } \\
 && \times
 \left[\frac{1}{2}\big\{k_1(t-t_1)+k_2(t-t_1)\big\}+i\varepsilon(t-t_1)\right]\rho_{10}(t_1).
 \nonumber
\end{eqnarray}
On the other hand, Eq.~\eqref{DYN-MAP-3} together with
Eq.~\eqref{G-DEF} yields:
\begin{eqnarray} \label{EQM-RHO10}
 \frac{d}{dt}\rho_{10}(t) = -\int_0^t dt_1 f(t-t_1) \rho_{10}(t_1),
\end{eqnarray}
where we have set the expansion parameter $\alpha$ equal to one
and we only have to remember that $f (t)$ is a quantity of second
order. Comparing Eqn.~\eqref{EQM-RHO10} with Eq.~\eqref{rho10} we
see that the expression within the square brackets of
\eqref{rho10} must be equal to $f(t-t_1)$, i.~e., we get the
conditions:
\begin{eqnarray}
 \varepsilon(\tau) &=& f_2(\tau),  \label{epsilon}\\
 k_1(\tau)+k_2(\tau) &=& 2f_1(\tau),\label{k1+k2}
\end{eqnarray}
where $f_1(\tau)$ and $f_2(\tau)$ denote the real and the
imaginary part of the correlation function:
\begin{equation}
\label{eq:1}
f(\tau) = f_1(\tau) + if_2(\tau).
\end{equation}
In order for Eq.~(\ref{rho11}) to reproduce the correct solution
\eqref{DYN-MAP-1} we have to choose $k_1(\tau)$ in such a way that
the solution of the equation
\begin{equation}
\label{eq:2}
 \frac{d}{dt} z(t) = -\int_0^t dt_1 k_1(t-t_1) z(t_1), \qquad
 z(0) = 1,
\end{equation}
is given by
\begin{equation}
\label{eq:3}
 z(t) = |G(t)|^2.
\end{equation}
Formulated in Laplace space this means that
\begin{equation} \label{k1hat}
 \hat{k}_1(u) = \frac{1-u\hat{z}(u)}{\hat{z}(u)}.
\end{equation}
Since the superoperator \eqref{NZ-kernel} preserves the
Hermiticity and the trace of the density matrix,
Eqs.~\eqref{DYN-MAP-1}-\eqref{DYN-MAP-4} follow directly from
Eqs.~\eqref{rho11} and \eqref{rho10}. Thus, we find that
Eq.~\eqref{NZ-kernel} represents the exact memory kernel of the
model for any given two-point correlation function. In fact, given
$f(\tau)$, the functions $\varepsilon(\tau)$, $k_1(\tau)$ and
$k_2(\tau)$ are uniquely determined by Eqs.~(\ref{epsilon}),
(\ref{k1+k2}) and (\ref{k1hat}). In view of this result the memory
kernel \eqref{NZ-kernel} can now be written in the form
\begin{eqnarray} \label{NZ-kernel-2}
 {\mathcal{K}}_{\rm NZ}(\tau) \rho &=&
 -if_2(\tau)[\sigma_+\sigma_-,\rho] - f_1(\tau)\left\{\sigma_+\sigma_-,\rho \right\}
 \nonumber \\
 && + k_1(\tau) \sigma_-\rho\sigma_+ \nonumber \\
 && + \left[2f_1(\tau)-k_1(\tau)\right]\sigma_+\sigma_-\rho\sigma_+\sigma_-,
\end{eqnarray}
which only involves the real and the imaginary part of the
correlation function and the function $k_1(\tau)$ which has to be
determined from Eq.~\eqref{k1hat}.

We note that the various coefficients in the memory kernel exhibit
a very different convergence behavior. In fact, we see that the
commutator and the anticommutator term in Eq.~\eqref{NZ-kernel-2}
come out exactly in second order in $\alpha$. It follows that the
equation of motion for the coherence $\rho_{10}$ [see
Eq.~\eqref{EQM-RHO10}] is already reproduced exactly within second
order, while the exact representation of the equation for the
population $\rho_{11}$ requires in general the inclusion of all
orders of the expansion. This non-uniform convergence behavior of
the elements of the density matrix has been observed also in
other, more complicated models \cite{Fischer2007a}, and seems to
be a typical feature of the perturbation expansion of the memory
kernel.

As will be shown below the relations \eqref{G-DEF} together with
\eqref{eq:2}-\eqref{k1hat} provide a direct perturbation approach
to the determination of the functions appearing in the memory
kernel Eq.~\eqref{NZ-kernel}, as an alternative to the standard
Nakajima-Zwanzig perturbation expansion. Moreover, this set of
equations allows in some cases to derive a closed analytical
expression for the memory kernel.

\subsection{Perturbation expansions of the memory kernel}\label{PERTURBATION}
Here we discuss two methods of expanding the exact memory kernel
with respect to the strength of the interaction Hamiltonian $H_I$.
The first expansion method relies on the expansion
Eq.~\eqref{expansion-G} for the solution of Eq.~\eqref{G-DEF},
which determines the dynamical map $\Phi (t)$ once the two-point
correlation function $f (t)$ of the model given by
Eq.~\eqref{eq:7} is specified.

Indeed as shown in Sec.\ref{sec:exact-memory-kernel} to obtain the
memory kernel  Eq.~\eqref{NZ-kernel} we only need to know the
function $k_1(t)$. A perturbative expression for the latter can be
easily obtained relying on the expansion Eq.~\eqref{expansion-G}
for $G(t)$, noting that thanks to Eq.~\eqref{eq:2} the Laplace
transform of $k_1(t)$ can be directly expressed through
Eq.~\eqref{k1hat} by means of the Laplace transform of the
function $z(t)=|G(t)|^2$. This procedure leads to the following
expansion
\begin{equation} \label{expansion-k1}
 k_1(t) = \sum_{n=0}^{\infty}  k_1^{(2n)}(t),
\end{equation}
as described in detail in Appendix~\ref{APP-A}, where the zero
order contribution is immediately seen to be zero.

Here we consider for the sake of simplicity only the second order
contribution, which is readily obtained. According to
Eq.~\eqref{eq:6} together with the initial condition $G (0)=1$ the
expression for $G(t)$ up to second order is given by
\begin{equation}
   \label{eq:9}
   G(t) \approx 1 - \int_0^t dt_1 \int_0^{t_1} dt_2 f(t_2),
\end{equation}
so that in the same approximation, recalling that the two-point
correlation function $f (t)$ is a quantity of second order, one has
\begin{equation}
   \label{eq:8}
    z(t) \approx 1 - 2\int_0^t dt_1 \int_0^{t_1} dt_2 f_1(t_2),
\end{equation}
where according to Eq.~\eqref{eq:1} $f_1 (t)$ denotes the real part of
the correlation function. The Laplace transform of this quantity is
now easily expressed in terms of the Laplace transform of the
correlation function according to
\begin{equation}
   \label{eq:10}
    \hat{z}(u) \approx \frac{u-2\hat{f_1}(u)}{u^2},
\end{equation}
and further exploiting Eq.~\eqref{k1hat} we find
\begin{equation}
   \label{eq:11}
   \hat k_1(u) \approx 2 \hat f_1(u).
\end{equation}
This immediately implies for the second-order contributions to the
kernel \eqref{NZ-kernel}
\begin{equation}
   \label{eq:12}
   k_1^{(2)}(t) = 2 f_1(\tau),
\end{equation}
and therefore due to Eq.~\eqref{k1+k2}
\begin{equation}
   \label{eq:13}
   k_2^{(2)}(\tau) =0.
\end{equation}
As shown in Appendix~\ref{APP-A} the fourth-order contribution reads
\begin{eqnarray}
\label{eq:20}
 k_{1}^{(4)}(t-t_1) &=& -2\Re \int^t_{t_1}dt_2 \int^{t_2}_{t_1}dt_3
 \left[ f(t-t_3) f(t_1-t_2) \right. \nonumber \\
 && \qquad\qquad + \left. f(t-t_1) f(t_3-t_2) \right],
\end{eqnarray}
so that $k_{2}^{(4)}=-k_{1}^{(4)}$. Indeed  Eq.~\eqref{k1+k2} generally
implies $k_{2}^{(2n)}=-k_{1}^{(2n)}$ for $n\geq 2$, therefore
Eq.~\eqref{expansion-k1} also provides an expansion for $k_2 (t)$.

The second expansion method is to employ the general
Nakajima-Zwanzig projection operator technique
\cite{Nakajima1958a,Zwanzig1960a} in which the memory kernel is
expressed in terms of the full propagator of the total system. The
details of this method for our model are presented in
Appendix~\ref{APP-B}, where it is shown that the projection
operator technique reproduces, as expected, the above results
obtained by the direct expansion of the coefficients in the memory
kernel.

\section{Discussion}
\label{sec:comp-time-local}

\subsection{Comparison of the time-convolutionless and the Nakajima-Zwanzig master equation}
\label{sec:time-conv-vers}

It is interesting to compare the time-convolutionless master
equation Eq.~\eqref{QMEQ} with the Nakajima-Zwanzig master
equation Eq.~\eqref{NZ-MASTEREQ}. For the considered model the
functions appearing in Eq.~\eqref{TCL-GEN} and
Eq.~\eqref{NZ-kernel} are given by Eq.~\eqref{eq:4} and
Eqs.~\eqref{epsilon}-\eqref{k1+k2} respectively.

Comparing Eqs.~\eqref{TCL-GEN} and \eqref{NZ-kernel} we see that
the superoperator structure of the memory kernel differs from that
of the time-convolutionless generator. In fact, the memory kernel
\eqref{NZ-kernel} contains the term proportional to $k_2(\tau)$
which involves the projection $\sigma_+\sigma_-=|1\rangle\langle
1|$ onto the excited state. Without such a term the equations
\eqref{rho11} and \eqref{rho10} for the population and the
coherence would be incompatible with the exact expressions
\eqref{DYN-MAP-1} and \eqref{DYN-MAP-3}. However, a term with this
structure is missing in the time-convolutionless generator
\eqref{TCL-GEN}. A further remarkable point is the fact that in
second order $k_2(\tau)=0$, according to Eqs.~\eqref{k1+k2} and
\eqref{eq:12}. This shows that the difference in the superoperator
structure of the memory kernel and the time-convolutionless
generator is visible only in higher orders of the perturbation
expansion.

The above discussion leads to some conclusions which are important
for the modelling of non-Markovian dynamics through
phenomenological master equations. In the Markovian limit our
model yields the following Lindblad generator ${\mathcal{L}}$
describing a quantum dynamical semigroup,
\begin{equation} \label{LINDBALD-GEN}
 {\mathcal{L}}\rho = -\frac{i}{2}S_M
 [\sigma_+\sigma_-,\rho]
 +\gamma_M\left[ \sigma_-\rho\sigma_+
 -\frac{1}{2}\left\{\sigma_+\sigma_-,\rho\right\} \right]
\end{equation}
with constant frequency shift $S_M$ and decay rate $\gamma_M \geq 0$.
Usually master equations of this form are derived by applying the
Markov approximation and second order perturbation theory (Born-Markov
approximation). A natural and widely-used non-Markovian generalization
is then obtained from this equation by keeping the structure of the
Lindblad generator ${\mathcal{L}}$ and by introducing a certain kernel
function $h(\tau)$ to arrive at a master equation of the form
\cite{Barnett2001a,Shabani2005a,Maniscalco2006a,Maniscalco2007a,Wilkie2009a}
\begin{equation} \label{ANSATZ}
 \frac{d}{dt} \rho(t) = \int_0^t dt_1 h(t-t_1) {\mathcal{L}} \rho(t_1).
\end{equation}
Although this equation is perfectly justified as a
phenomenological ansatz, it does in general not represent the
correct structure of the memory kernel of the underlying
microscopic model. In fact, we see that even for the simple model
studied here the true memory kernel \eqref{NZ-kernel} is not of
the form of Eq.~\eqref{ANSATZ}, but involves additional terms that
are absent in the Markovian Lindblad generator.  Indeed it is
rather given by a linear combination of terms of the form
Eq.~\eqref{ANSATZ}, where going beyond the Born approximation
besides the Markovian Lindblad generator other operator structures
appear, which are still in Lindblad form but with different
Lindblad operators.  This observation seems to be of particular
relevance for the analysis of the positivity and the complete
positivity of the dynamical maps obtained from phenomenological
equations of motion.

\subsection{Example}
\label{sec:example} These considerations can be nicely illustrated
considering the example of an exponential correlation function,
corresponding to a Lorentzian spectral density \cite{Breuer2007}
\begin{equation} \label{exp-corr}
  f(\tau) = \frac{1}{2} \gamma_0 \lambda e^{-\lambda |\tau|},
\end{equation}
where the parameters $\gamma_0$ and $\lambda$ are real and
positive.
For this case both time-convolutionless generator and Nakajima-Zwanzig
kernel can be exactly calculated. Indeed by means of Eq.~\eqref{G-DEF}
one obtains for the function
$G(t)$ the expression:
\begin{equation}
\label{eq:16}
  G(t) = e^{-\lambda t/2} \left[ \cosh \left(
  \frac{\lambda t}{2} \delta\right) + \frac{1}{\delta} \sinh \left(
  \frac{\lambda t}{2} \delta \right)
  \right],
\end{equation}
where $\delta=\sqrt{1-2\gamma_0/\lambda}$.
Note that this function is always real. Furthermore it stays positive
for any time $t$ in the weak coupling regime $\gamma_0<\lambda/2$,
while for strong coupling $\gamma_0>\lambda/2$ the parameter $\delta$
becomes purely imaginary and the function
$G(t)$ starts to oscillate. In particular it goes through zero for the
first time when $t$ is equal to the smallest positive solution of
\begin{equation}
   \label{eq:15}
t=\frac{2}{\lambda \hat\delta}\left(n\pi - \arctan \hat\delta \right),
\end{equation}
where $\hat\delta=\sqrt{2\gamma_0/\lambda-1}$ and $n\in
\mathbb{N}$. Building on Eq.~\eqref{eq:16} one can obtain the
exact expressions for the functions $\gamma (t)$ and $S (t)$
appearing in the time-convolutionless generator, given by $S(t)=0$
and
\begin{equation}
   \label{eq:17}
   \gamma (t)={2\gamma_0}\frac{\sinh \left(
        \frac{\lambda t}{2} \delta \right)}{\delta \cosh \left(
  \frac{\lambda t}{2} \delta\right) + \sinh \left(
  \frac{\lambda t}{2} \delta \right)}.
\end{equation}
In order to obtain the Nakajima-Zwanzig kernel one considers the
Laplace transform of the function $z(t)=|G(t)|^2$ which is found to be:
\begin{equation}
 \hat{z}(u) = \frac{(u+\lambda)(u+2\lambda)
 +\gamma_0\lambda}{(u+\lambda) \left[(u+\lambda)^2 - \lambda^2
    +2\gamma_0 \lambda \right]},
\end{equation}
so that according to Eq.~\eqref{k1hat} one has:
\begin{eqnarray}
  \hat{k_1}(u) =  \gamma_0 \lambda \frac{u+2\lambda}{(u+\lambda)(u+2\lambda)
  + \gamma_0 \lambda}.
\end{eqnarray}
Transforming back to the time domain we finally get
\begin{equation} \label{k1-exp-corr}
 k_1(t) = \gamma_0 \lambda e^{- 3\lambda t /2}
  \left[ \cosh \left( \frac{\lambda t}{2}\delta'  \right) + \frac{1}{\delta'} \sinh
  \left( \frac{\lambda t}{2}\delta' \right) \right],
\end{equation}
where $\delta'=\sqrt{1-4\gamma_0/\lambda}$. Substituting this
result into Eq.~\eqref{NZ-kernel-2} we find the exact memory
kernel for the case of an exponential correlation function.

The exact expressions Eq.~\eqref{eq:17} and Eq.~\eqref{k1-exp-corr}
already allow for an important comparison. While the function on the right-hand
side of Eq.~\eqref{k1-exp-corr} represents an analytic function of
$\gamma_0$ (remember that $\gamma_0$ is a quantity of
second order in the expansion parameter $\alpha$), so that the
Nakajima-Zwanzig memory kernel has an
infinite radius of convergence, the same does not hold true for
the time-convolutionless generator. Indeed the time-convolutionless
expansion
breaks down in the strong coupling regime $\gamma_0>\lambda/2$ when
the function $G (t)$ given in Eq.~\eqref{eq:16} goes through zero,
corresponding to the divergence of the decay rate $\gamma (t)$ given
in Eq.~\eqref{eq:17} and obtained from the relation Eq.~\eqref{eq:4}.

Considering an expansion in $\gamma_0$
of the function $k_1(t)$ given by Eq.~\eqref{k1-exp-corr} which fixes
the memory kernel, due to the fact that the correlation function
Eq.~\eqref{exp-corr} is real one obtains
\begin{eqnarray} \label{kernel-second-order}
 {\mathcal{K}}^{(2)}_{\rm NZ}(\tau) \rho =
 2f(\tau) \left[ \sigma_-\rho\sigma_+
 - \frac{1}{2} \left\{ \sigma_+\sigma_-,\rho \right\} \right],
\end{eqnarray}
so that up to second-order the corresponding master equation is indeed of the form of
Eq.~\eqref{ANSATZ} with the exponential kernel function
$h(t)=2f(t)$. However, in fourth order further terms appear which
are not present in \eqref{ANSATZ}:
\begin{eqnarray} \label{kernel-fourth-order}
 {\mathcal{K}}^{(4)}_{\rm NZ}(\tau) \rho =
 k_1^{(4)}(\tau) \left[ \sigma_-\rho\sigma_+ -
 \sigma_+\sigma_-\rho\sigma_+\sigma_-\right],
\end{eqnarray}
where
\begin{equation}
 k_1^{(4)}(\tau) = \gamma_0^2 \left[e^{-\lambda \tau}(1-\lambda \tau)-e^{-2\lambda \tau} \right].
\end{equation}
As shown in \cite{Breuer2009a} this implies in particular that if
one truncates the expansion to first order in $\gamma_0$ the
complete positivity (and even the positivity) of the resulting
dynamical map is violated for strong couplings in the
Nakajima-Zwanzig case. On the contrary the second-order
time-convolutionless master equation always guarantees complete
positivity, as can be seen considering the second-order
approximation for Eq.~\eqref{eq:17} given by
\begin{equation}
   \label{eq:18}
   \gamma^{(2)} (t)=\gamma_0\left (1-e^{-\lambda t}\right).
\end{equation}

\section{Conclusions}\label{sec:disc-concl}

We have constructed the exact Nakajima-Zwanzig memory kernel for a
specific model describing the decay of a two-level system into a
reservoir of field modes which is initially in the vacuum state.
The construction of the memory kernel is based on the analytical
solution of the Schr\"odinger equation within the Hilbert space
sector describing states with zero or one excitation, and is valid for
a generic spectral density. Since the
dynamical map giving the reduced system dynamics of the two-state
system is known exactly, there is of course in principle no reason
to resort to any kind of master equation in order to determine the
dynamical behavior of the system. However, the present results
lead to several important implications which are relevant for more
realistic physical systems and their microscopic or
phenomenological modelling, where analytical results cannot be
obtained. Indeed for this model both time-convolutionless
generator and Nakajima-Zwanzig kernel can be exactly expressed in
terms of functions for which perturbative expansions are given,
together with the exact solution for a reservoir with an
exponential correlation function, corresponding to a Lorentzian
spectral density. This allows for a detailed comparison of the two
approaches expressing the dynamics in terms of a time-local and
integrodifferential master equation respectively. It turns out
that contrary to what is often expected the Nakajima-Zwanzig
master equation is not simply obtained by convolution of the
Lindblad operator appearing in the non-Markovian case with a
suitable kernel. It actually has a different operator structure,
emerging when considering higher perturbative orders. Furthermore
the exact analytical result obtained for a Lorentzian spectral
density shows the different convergence behavior of the two
approaches. While the Nakajima-Zwanzig kernel is an analytic
function of the coupling strength, providing a well-defined master
equation at any time, the time-convolutionless generator breaks
down at finite time in the strong coupling regime, thus failing to
reproduce the asymptotic behavior.

 \begin{acknowledgments}
This work was partially supported by MIUR under PRIN2008.
 \end{acknowledgments}

\begin{widetext}

\appendix

\section{}\label{APP-A}

In this Appendix we shall consider how to obtain a perturbative expansion for
the function $k_1 (\tau)$ which according to
Eqs.~(\ref{NZ-kernel}) and (\ref{epsilon})-(\ref{k1+k2}) determines the memory
kernel in the Nakajima-Zwanzig master equation, as a function of the two-point
correlation function $f (t)$ of the reservoir. To this end one
considers the solution of Eq.~(\ref{G-DEF}) which is of the form
Eq.~(\ref{expansion-G}) with $G \left( 0 \right) = 1$ and $G^{\left( 2 n
\right)} (t)$ explicitly given by
\begin{eqnarray}\label{eq:22}
  G^{\left( 2 n \right)} (t) & = & \left( - \right)^n \int^t_0
  dt_1 \int^{t_1}_0 dt_2 \cdots \int^{t_{2 n - 1}}_0 dt_{2 n} \prod^n_{i = 1}
  \bignone f \left( t_{2 i - 1} - t_{2 i} \right)
\end{eqnarray}
so that
\begin{eqnarray}\label{eq:23}
  z (t) = \left| G (t) \right|^2 & = & 1 + 2
  \Re \sum^{\infty}_{n = 1} \left( - \right)^n \int^t_0 dt_1
  \int^{t_1}_0 dt_2 \cdots \int^{t_{2 n - 1}}_0 dt_{2 n} \prod^n_{i = 1}
  \bignone f \left( t_{2 i - 1} - t_{2 i} \right) \\
  &  & + \left| \sum^{\infty}_{n = 1} \left( - \right)^n \int^t_0 dt_1
  \int^{t_1}_0 dt_2 \cdots \int^{t_{2 n - 1}}_0 dt_{2 n} \prod^n_{i = 1}
  \bignone f \left( t_{2 i - 1} - t_{2 i} \right) \right|^2 . \nonumber
\end{eqnarray}
Considering terms up to fourth order in the expansion parameter one
has
\begin{eqnarray}\label{eq:24}
  z (t) & = & 1 - 2 \Re \int^t_0 dt_1 \int^{t_1}_0 dt_2 f
  \left( t_2 \right)\\
  &  & + 2 \Re \int^t_0 dt_1 \int^{t_1}_0 dt_2 \int^{t_2}_0 dt_3 \int^{t_4}_0
  dt_4 f \left( t_1 - t_2 \right) f \left( t_3 - t_4 \right) \nonumber\\
  &  & + \left| \int^t_0 dt_1 \int^{t_1}_0 dt_2 f \left( t_2 \right)
  \right|^2 + \ldots, \nonumber
\end{eqnarray}
and denoting real and imaginary parts of $f (t)$ as in Eq.~\eqref{eq:1} also
\begin{eqnarray}\label{eq:25}
  z (t) & = & 1 - 2 \int^t_0 dt_1 \int^{t_1}_0 dt_2 f_1 \left(
  t_2 \right)\\
  &  & + 2 \int^t_0 dt_1 \int^{t_1}_0 dt_2 \int^{t_2}_0 dt_3 \int^{t_3}_0
  dt_4 \left[ f_1 \left( t_1 - t_2 \right) f_1 \left( t_3 - t_4 \right) - f_2
  \left( t_1 - t_2 \right) f_2 \left( t_3 - t_4 \right) \right] \nonumber\\
  &  & + \left| \int^t_0 dt_1 \int^{t_1}_0 dt_2 f_1 \left( t_2 \right)
  \right|^2 + \left| \int^t_0 dt_1 \int^{t_1}_0 dt_2 f_2 \left( t_2 \right)
  \right|^2 + \ldots . \nonumber
\end{eqnarray}
Introducing the functions
\begin{eqnarray}\label{eq:26}
  h_i (t) & = & \int^t_0 dt_1 \int^{t_1}_0 dt_2 f_i \left( t_2
  \right) \hspace{2em} i = 1, 2
\end{eqnarray}
one obtains for the Laplace transform of $z (t)$
\begin{eqnarray}\label{eq:27}
  \hat{z} (u) & = & \frac{u - 2 u \widehat{f_1} \left( u
  \right)}{u^2} + \frac{2}{u^3} \left( \widehat{f_1}^2 (u) -
  \widehat{f_2}^2 (u) \right) + \widehat{h^2_1} (u)
  + \widehat{h^2_2} (u) + \ldots,
\end{eqnarray}
and thanks to Eq.~\eqref{k1hat}
\begin{eqnarray}\label{eq:k12l}
  \hat{k}_1 (u) & = & 2 \widehat{f_1} (u) +
  \frac{2}{u^{}} \left( \widehat{f_1}^2 (u) + \widehat{f_2}^2
  (u) \right) - u^2 \left( \widehat{h^2_1} (u) +
  \widehat{h^2_2} (u) \right) + \ldots .
\end{eqnarray}
Using now the fact that the functions $h_i$ are equal to zero together with
their derivatives at $t=0$ so that
\begin{eqnarray}\label{eq:28}
  \widehat{\frac{\mathd^2}{dt^2}h^2_i}  (u) & = & u^2
  \widehat{h^2_i} (u),
\end{eqnarray}
one has
\begin{eqnarray}\label{eq:29}
  k_1^{} (\tau) & = & 2 f_1 (\tau) + 2 \int^{\tau}_0
  dt_1 \int^{t_1}_0 dt_2 f_1 \left( t_1 - t_2 \right) f_1 \left( t_2 \right) +
  2 \int^{\tau}_0 dt_1 \int^{t_1}_0 dt_2 f_2 \left( t_1 - t_2 \right) f_2
  \left( t_2 \right)\\
  &  & - 2 \left| \int^{\tau}_0 dt_1 f_1 \left( t_1 \right) \right|^2 - 2
  \left| \int^{\tau}_0 dt_1 f_2 \left( t_1 \right) \right|^2 \nonumber\\
  &  & - 2 f_1 (\tau) \int^{\tau}_0 dt_1 \int^{t_1}_0 dt_2 f_1
  \left( t_1 - t_2 \right) - 2 f_2 (\tau) \int^{\tau}_0 dt_1
  \int^{t_1}_0 dt_2 f_2 \left( t_1 - t_2 \right) + \ldots . \nonumber
\end{eqnarray}
We now exploit the identity
\begin{eqnarray}
\label{eq:21}
  \int^{\tau}_0 dt_2  \int_0^{t_2} dt_3 f \left( t_2 - t_3 \right) f
  \left( t_3 \right) - \left| \int^{\tau}_0 dt_2 f_1 \left( t_2 \right)
  \right|^2 + \int^{\tau}_0 dt_2 \int_0^{t_2} dt_3 f \left( \tau - t_3
  \right) f \left( t_2 \right) & = & 0
\end{eqnarray}
which can be checked noting that the function of $t$ defined by the left-hand
side of Eq.~\eqref{eq:21} has vanishing derivative and is equal to zero for $t =
0$. We are thus left with
\begin{eqnarray}\label{eq:k12tbis}
  k_1^{} (\tau) & = & 2 f_1 (\tau) - 2 \int^{\tau}_0
  dt_2 \int^{t_2}_0 dt_3 \left[ f_1 \left( \tau - t_3 \right) f_1 \left( t_2
  \right) + f_1 (\tau) f_1 \left( t_2 - t_3 \right) \right]
  \\
  &  & - 2 \int^{\tau}_0 dt_2 \int^{t_2}_0 dt_3 \left[ f_2 \left( \tau - t_3
  \right) f_2 \left( t_2 \right) + f_2 (\tau) f_2 \left( t_2 -
  t_3 \right) \right] + \ldots \nonumber\\
  &  &  \nonumber
\end{eqnarray}
and thanks to the fact that real and imaginary parts of $f (t)$
are even and odd respectively
\begin{eqnarray}\label{eq:30}
  k_1^{} (\tau) & = & 2 f_1 (\tau) - 2 \int^{\tau}_0
  dt_2 \int^{t_2}_0 dt_3 \left[ f_1 \left( \tau - t_3 \right) f_1 \left( - t_2
  \right) - f_2 \left( \tau - t_3 \right) f_2 \left( - t_2 \right)
   \right.\\
  &  & \left. + f_1 (\tau) f_1 \left( t_3 - t_2 \right) - f_2
  (\tau) f_2 \left( t_3 - t_2 \right) \right] + \ldots \nonumber\\
  & = & 2  f_1 (\tau) - 2 \Re \int^{\tau}_0 dt_2 \int^{t_2}_0
  dt_3 \left[ f \left( \tau - t_3 \right) f \left( - t_2 \right) + f \left(
  \tau \right) f \left( t_3 - t_2 \right) \right] + \ldots \nonumber
\end{eqnarray}
Upon the change of variables $t_2 \rightarrow t_2 - t_1$, $t_3 \rightarrow
t_3 - t_1$ one has for the second and fourth order contribution to $k_1^{}
(\tau)$:
\begin{eqnarray}\label{eq:31}
  k_1^{\left( 2 \right)} \left( t - t_1 \right) & = & 2  f_1 \left( t -
  t_1 \right)  \\
  k_1^{\left( 4 \right)} \left( t - t_1 \right) & = & - 2 \Re \int^t_{t_1}
  dt_2 \int^{t_2}_{t_1} dt_3 \left[ f \left( t - t_3 \right) f \left( t_{1
  \,} - t_2 \right) + f \left( t - t_1 \right) f \left( t_3 - t_2
  \right) \right] .
\end{eqnarray}

\section{}\label{APP-B}

Here we derive the contributions up to fourth order to the memory
kernel Eq.~\eqref{NZ-kernel} employing the standard
Nakajima-Zwanzig projection operator technique. Since the initial
state of system and bath is of the factorized form
Eq.~\eqref{INIT-PROD} we can employ the standard projection
operator
\begin{eqnarray}\label{eq:32}
  \mathcal{P} w & = & \tmop{Tr}_E \left( w \right) \otimes \rho_E,
\end{eqnarray}
where $w$ is a state of system plus environment and $\rho_E$ denotes
the vacuum state of the reservoir. This projection operator is the
same used to obtain Eq.~\eqref{cumulants} and for it the initial state
Eq.~\eqref{INIT-PROD} is indeed an eigenoperator. Introducing further
the superoperators
\begin{eqnarray}\label{eq:33}
  \mathcal{L} (t) \rho (t) & = & - i \left[ H_I
  (t), \rho (t) \right]
\end{eqnarray}
with $H_I (t)$ as in Eq.~\eqref{eq:hi}, and
\begin{eqnarray}\label{eq:34}
  \mathcal{G} \left( t, t_1 \right) & = & \mathcal{T} \exp \left( \int^t_{t_1}
  \mathd s \mathcal{QL} \left( s \right) \right)^{\bignone}
\end{eqnarray}
where $\mathcal{T}$ denotes time ordering and $\mathcal{Q} = 1 - \mathcal{P}$,
the Nakajima-Zwanzig memory kernel appearing in Eq.~\eqref{NZ-MASTEREQ} is given by
\begin{eqnarray}\label{eq:35}
  \mathcal{K}_{\tmop{NZ}} \left( t - t_1 \right) \rho \left( t_1 \right) & = &
  \tmop{Tr}_E \left( \mathcal{L} (t) \mathcal{G} \left( t, t_1
  \right) \mathcal{Q} \mathcal{L} \left( t_1 \right) \mathcal{} \rho \left(
  t_1 \right) \otimes \rho_E \right) .
\end{eqnarray}
Noting that for this model $\mathcal{P} \mathcal{L} \left( t_1 \right) \ldots
\mathcal{L} \left( t_{2 n + 1} \right) \mathcal{P} = 0$ one has
\begin{eqnarray}\label{eq:36}
  \mathcal{K}^{}_{\tmop{NZ}} \left( t - t_1 \right) \rho \left( t_1 \right) &
  = & \tmop{Tr}_E \left( \mathcal{L} (t) \mathcal{L} \left( t_1
  \right) \mathcal{} \rho \left( t_1 \right) \otimes \rho_E \right)
  \\
  &  & + \int^t_{t_1} dt_2 \int^{t_2}_{t_1} dt_3 \left[ \tmop{Tr}_E \left(
  \mathcal{L} (t) \mathcal{L} \left( t_2 \right) \mathcal{L}
  \left( t_3 \right) \mathcal{L} \left( t_1 \right) \rho \left( t_1 \right)
  \otimes \rho_E \right) \right. \nonumber\\
  &  & \left. - \tmop{Tr}_E \left( \mathcal{L} (t) \mathcal{L}
  \left( t_2 \right) \mathcal{P} \mathcal{L} \left( t_3 \right) \mathcal{L}
  \left( t_1 \right) \rho \left( t_1 \right) \otimes \rho_E \right) \right] +
  \ldots \nonumber
\end{eqnarray}
Using Eq.~\eqref{eq:hi} and Eq.~\eqref{eq:7} one readily obtains
\begin{eqnarray}\label{eq:37}
  \tmop{Tr}_E \left( \mathcal{L} (t) \mathcal{L} \left( t_1
  \right) \mathcal{} \rho \left( t_1 \right) \otimes \rho_E \right) & = &
  \left( - i \right)^2 \left[ f \left( t - t_1 \right) \sigma_+ \sigma_-
  \rho_{} \left( t_1 \right) - f \left( t_1 - t \right) \sigma_- \rho_{}
  \left( t_1 \right) \sigma_+ \right. \\
  &  & - f \left( t - t_1 \right) \sigma_- \rho_{} \left( t_1 \right)
  \sigma_+ + f \left( t_1 - t \right) \rho_{} \left( t_1 \right) \sigma_+
  \sigma_-], \nonumber
\end{eqnarray}
so that the second order contribution is given by
\begin{eqnarray}\label{eq:38}
  \mathcal{K}^{\left( 2 \right)}_{\tmop{NZ}} (\tau) \rho & = & -
  i f_2 (\tau) \left[ \sigma_+ \sigma_-, \rho \right]
\\
  &  & + 2 f_1 (\tau) \left[ \sigma_- \rho \sigma_+ -
  \frac{1}{2} \left\{ \sigma_+ \sigma_-, \rho \right\} \right], \nonumber
\end{eqnarray}
which due to Eq~.\eqref{NZ-kernel-2} confirms the result Eq.~\eqref{eq:12}. Setting
\begin{eqnarray}\label{eq:39}
  I_1 \left( t, t_2, t_3, t_1 \right) \rho \left( t_1 \right) & = &
  \tmop{Tr}_B \left\{ \mathcal{L} (t) \mathcal{L} \left( t_2
  \right) \mathcal{L} \left( t_3 \right) \mathcal{L} \left( t_1 \right)
  \rho \left( t_1 \right) \otimes \rho_E \right\}
\end{eqnarray}
and
\begin{eqnarray}\label{eq:40}
  I_2 \left( t, t_2, t_3, t_1 \right) \rho \left( t_1 \right) & = &
  \tmop{Tr}_B \left\{ \mathcal{}  \mathcal{L} (t) \mathcal{L}
  \left( t_2 \right) \mathcal{P} \mathcal{L} \left( t_3 \right) \mathcal{L}
  \left( t_1 \right)  \rho \left( t_1 \right) \otimes \rho_E
   \right\}
\end{eqnarray}
a lengthy but straightforward calculation leads to the results
\begin{eqnarray}\label{eq:41}
  I_2 \left( t, t_2, t_3, t_1 \right) \rho \left( t_1 \right) & = & f
  \left( t - t_2 \right) f \left( t_3 - t_1 \right) \sigma_+ \sigma_- \rho
  \left( t_1 \right) + f \left( t_2 - t \right) f \left( t_1 - t_3
  \right) \rho \left( t_1 \right) \sigma_+ \sigma_- \\
  &  & + 2 \Re \left[ f \left( t - t_2 \right) f \left( t_1 - t_3 \right)
  \right] \sigma_+ \sigma_- \rho \left( t_1 \right) \sigma_+ \sigma_-
  \nonumber\\
  &  & - 4 f_1 \left( t - t_2 \right) f_1 \left( t_1 - t_3 \right)
  \sigma_- \rho \left( t_1 \right) \sigma_+ \nonumber
\end{eqnarray}
and
\begin{eqnarray}\label{eq:42}
  I_1 \left( t, t_2, t_3, t_1 \right) \rho \left( t_1 \right) & = & I_2 \left(
  t, t_2, t_3, t_1 \right) \rho \left( t_1 \right)\\
  &  & - 2 \Re \left[ f \left( t - t_3 \right) f \left( t_1 - t_2 \right)
  + f \left( t - t_1 \right) f \left( t_3 - t_2 \right) \right] \sigma_-
  \rho \left( t_1 \right) \sigma_+ \nonumber\\
  &  & + 2 \Re \left[ f \left( t - t_3 \right) f \left( t_1 - t_2 \right)
  + f \left( t - t_1 \right) f \left( t_3 - t_2 \right) \right] \sigma_+
  \sigma_- \rho \left( t_1 \right) \sigma_+ \sigma_- . \nonumber
\end{eqnarray}
One thus have for the fourth order expression
\begin{eqnarray}\label{eq:nz4}
  \mathcal{K}^{\left( 4 \right)}_{\tmop{NZ}} \left( t - t_1 \right) \rho & = &
  - 2 \Re \int^t_{t_1} dt_2 \int^{t_2}_{t_1} dt_3 \left[ f \left( t - t_3
  \right) f \left( t_1 - t_2 \right) + f \left( t - t_1 \right) f \left(
  t_3 - t_2 \right) \right]\\
  &  & \times \left[ \sigma_- \rho \sigma_+ - \sigma_+ \sigma_- \rho \sigma_+
  \sigma_- \right], \nonumber
\end{eqnarray}
which according to Eq.~\eqref{NZ-kernel-2} confirms Eq.~\eqref{eq:20}.

\end{widetext}


\begin{thebibliography}{33}

\bibitem{Alicki2007}
R.~Alicki and K.~Lendi, \emph{Quantum Dynamical Semigroups and Applications},
  Vol. 717 of \emph{Lecture Notes in Physics}, 2nd~edn. (Springer, Berlin,
  2007)

\bibitem{Breuer2007}
H.-P. Breuer and F.~Petruccione, \emph{The Theory of Open Quantum Systems}
  (Oxford University Press, Oxford, 2007)

\bibitem{Weiss2008}
U.~Weiss, \emph{{Quantum Dissipative Systems}}, 3rd~edn. (World Scientific,
  Singapore, 2008)

\bibitem{Gorini1976a}
V.~Gorini, A.~Kossakowski, and E.~C.~G. Sudarshan, J.~Math. Phys. \textbf{17},
  821 (1976)

\bibitem{Lindblad1976a}
G.~Lindblad, Comm. Math. Phys. \textbf{48}, 119 (1976)

\bibitem{Breuer2007a}
H.-P. Breuer, Phys. Rev.~A \textbf{75}, 022103 (2007)

\bibitem{Breuer2006a}
H.-P. Breuer, J.~Gemmer, and M.~Michel, Phys. Rev.~E \textbf{73}, 016139 (2006)

\bibitem{Breuer2008a}
H.-P. Breuer and B.~Vacchini, Phys. Rev. Lett. \textbf{101}, 140402 (2008)

\bibitem{Breuer2009a}
H.-P. Breuer and B.~Vacchini, Phys. Rev.~E \textbf{79}, 041147 (2009)

\bibitem{Budini2005a}
A.~A. Budini, Phys. Rev.~E \textbf{72}, 056106 (2005)

\bibitem{Budini2006a}
A.~A. Budini, Phys. Rev.~A \textbf{74}, 053815 (2006)

\bibitem{Budini2004a}
A.~A. Budini, Phys. Rev.~A \textbf{69}, 042107 (2004)

\bibitem{Budini2005b}
A.~A. Budini and H.~Schomerus, J.~Phys.~A: Math. Gen. \textbf{38}, 9251 (2005)

\bibitem{Chruscinski-xxx}
D.~Chruscinski and A.~Kossakowski (2009), eprint arXiv:0912.1259v2

\bibitem{Chruscinski-xxx_1}
D.~Chruscinski, A.~Kossakowski, and S.~Pascazio (2009), eprint
  arXiv:0906.5122v2

\bibitem{Ferraro2008a}
E.~Ferraro, H.-P. Breuer, A.~Napoli, M.~A. Jivulescu, and A.~Messina, Phys.
  Rev.~B \textbf{78}, 064309 (2008)

\bibitem{Kossakowski2008a}
A.~Kossakowski and R.~Rebolledo, Open Syst. Inf. Dyn. \textbf{15}, 135 (2008)

\bibitem{Kossakowski2009a}
A.~Kossakowski and R.~Rebolledo, Open Syst. Inf. Dyn. \textbf{16}, 259 (2009)

\bibitem{Krovi2007a}
H.~Krovi, O.~Oreshkov, M.~Ryazanov, and D.~A. Lidar, Phys. Rev.~A \textbf{76},
  052117 (2007)

\bibitem{Piilo2008a}
J.~Piilo, S.~Maniscalco, K.~Harkonen, and K.-A. Suominen, Phys. Rev. Lett.
  \textbf{100}, 180402 (2008)

\bibitem{Vacchini2008a}
B.~Vacchini, Phys. Rev.~A \textbf{78}, 022112 (2008)

\bibitem{Breuer2009b}
H.-P. Breuer, E.-M. Laine, and J.~Piilo, Phys. Rev. Lett. \textbf{103}, 210401
  (2009)

\bibitem{Rivas-xxx}
A.~Rivas, S.~F. Huelga, and M.~B. Plenio (2009), eprint arXiv:0911.4270v1

\bibitem{Barnett2001a}
S.~M. Barnett and S.~Stenholm, Phys. Rev.~A \textbf{64}, 033808 (2001)

\bibitem{Maniscalco2007a}
S.~Maniscalco, Phys. Rev.~A \textbf{75}, 062103 (2007)

\bibitem{Maniscalco2006a}
S.~Maniscalco and F.~Petruccione, Phys. Rev.~A \textbf{73}, 012111 (2006)

\bibitem{Shabani2005a}
A.~Shabani and D.~A. Lidar, Phys. Rev.~A \textbf{71}, 020101 (2005)

\bibitem{Wilkie2009a}
J.~Wilkie and Y.~M. Wong, J.~Phys.~A: Math. Gen. \textbf{42}, 015006 (2009)

\bibitem{Daffer2004a}
S.~Daffer, K.~W{\'o}dkiewicz, J.~D. Cresser, and J.~K. McIver, Phys. Rev.~A
  \textbf{70}, 010304 (2004)

\bibitem{Weisskopf1930a}
V.~Weisskopf and E.~Wigner, Z.~Physik \textbf{63}, 54 (1930)

\bibitem{Fischer2007a}
J.~Fischer and H.-P. Breuer, Phys. Rev.~A \textbf{76}, 052119 (2007)

\bibitem{Nakajima1958a}
S.~Nakajima, Progr. Theor. Phys. \textbf{20}, 948 (1958)

\bibitem{Zwanzig1960a}
R.~Zwanzig, J.~Chem. Phys. \textbf{33}, 1338 (1960)

\end{thebibliography}


\end{document}